\documentclass{kluwer}    
\usepackage{psfig,latexsym,longtable,lscape,supertabular}
\newdisplay{guess}{Conjecture}

\begin{document}                                                                                   
\begin{article}
\begin{opening}         
\title{X-ray halos in galaxies
		} 
\author{Ginevra \surname{Trinchieri}}  
\runningauthor{Ginevra Trinchieri}
\runningtitle{X-ray halos in galaxies}
\institute{INAF-Osservatorio Astronomico di Brera}
\date{\today}

\begin{abstract}
A hot phase of the interstellar medium has now been detected and studied
in several objects through its X-ray emission.  A proper assessment of
its characteristics  is relevant for our understanding of several aspects
of galaxy properties, from the large scale distribution of matter to the
stellar and galaxian evolution, to the dynamics of systems and to
the feeding of a central black hole. I will briefly summarize
our current understanding of some of the main issues related to the hot gaseous
component in galaxies that are fast evolving given the ever more
striking and interesting details provided by the
X-ray satellites currently operating. 
I hope to convince you that the  X-ray characteristics
of the hot gas are quite complex, both in
morphology and spectra, in a wide range of objects,  which should
promote greater efforts in understanding the role played
by this component in all galaxies.
\end{abstract}

\end{opening}           

\section{Why is hot gas important?}  

Because of the implications and relations that a hot phase of the
interstellar/intergalactic medium has with several other galaxy-related
issues, understanding its properties has become more
relevant and intriguing. As just a few examples, gas is closely related
to: 

\smallskip\noindent
$\bullet$ galaxy structure.  Gas can be traced at large
galactic radii and has been used to study and measure the total
gravitational mass in early type galaxies (e.g. Fabricant \& Gorenstein
1983) \\
$\bullet$ evolution of stars, of 
galaxies and of larger systems.  Metal enriched material that is formed
in stars gives a measure of 
stellar evolution; metal gradients indicate how far (how?) metals have traveled
from their formation sites; metal
concentrations in specific locations could indicate local stripping
phenomena \\
$\bullet$ the dynamics of systems.  At high energies 
different phenomena such as stripping and shocks could be
discovered and studied \\
$\bullet$ active nuclei, who are nourished by gas accretion.  When hot, the
gas could indeed starve the nucleus rather than feed it, and explain 
the lack of activity or periodic cycles \\

\smallskip\noindent The main question is then to find the best candidates to
efficiently study the properties of this phase of the
InterStellarMedium (ISM).

\begin{figure}
\unitlength1.0cm
\begin{picture}(11,6.0)
\thicklines
\put(0,-0.1){
\psfig{file=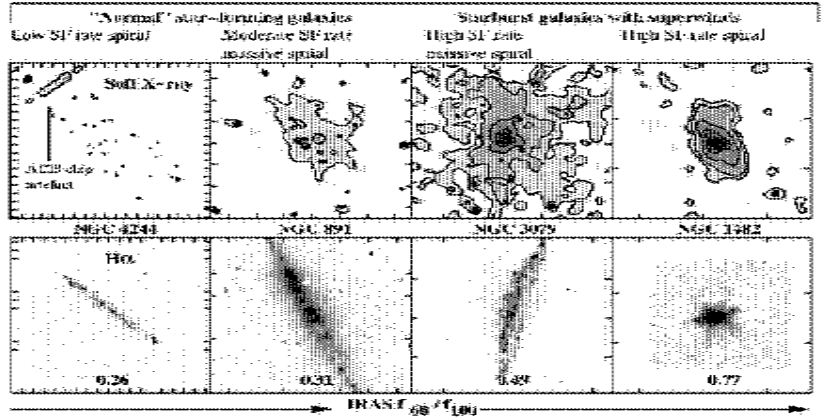,width=12cm,height=6cm}
}
\end{picture}
\caption[]{From \citeauthor{stricketal} (\citeyear{stricketal}). Top panels: X-ray images in the
0.3-2.0 keV energy band from Chandra data for 4 galaxies from normal to
actively forming stars.  Lower panels:  The same galaxies in the
continuum-subtracted H$\alpha$+[N{\sc ii}] emission. The arrow at the
bottom indicates the increasing IRAS $f_{60}/f_{100}$ ratio, which
is used as a measure of the star formation (SF) intensity.  
The increasingly strong soft
X-ray emission associated with increasing SF activity is evident. 
}
\label{strick}
\end{figure}

\section{Where is the hot gas?}

There is very little [if any] diffuse emission in normal spiral
galaxies.  Gas appears to be associated  mostly with:
\begin{enumerate}
\item \underline{\bf star formation}. 
A figure from a talk by \citeauthor{stricketal}\citeyear{stricketal}
provides an excellent example and confirmation of this 
association.  The X-ray images
reveal more and more extended soft diffuse emission from hot gas as the
line emission and star formation activity  become 
more prominent (see Fig.\ref{strick}). 
\item \underline{\bf gravity}. 
Large quantities of hot gas have been discovered in early type galaxies
since $Einstein$ observations, and are associated mostly with 
galaxies at the center of small groups.  A long
standing issue of whether the large scale gas should be
associated with the galaxy
or with the group is alive and unresolved as yet, but it is beyond the
scope of this talk and it will be mostly [unjustly!] ignored here. 
\end{enumerate}

Smaller quantities of gas have recently  been found associated with
spiral bulges (e.g. M31, NGC~1291) and low luminosity early type galaxies
(e.g. NGC~4697), but these
are still hard to quantify and characterize.  If this is a property of
the whole class, a new and interesting field will open up and bring new
issues to study and resolve. 

It is evident that the characteristics of the gas might not be the same
in both cases above, considering the clearly different origins of the
two phenomena.  It is however likely that different systems will be
dominated by only one of the two components, thereby  allowing us a
relatively clean study of both phenomena separately.

\begin{figure*}
\unitlength1.0cm
\begin{picture}(12,6.0)
\thicklines
\put(-0.3,0){
\begin{picture}(6.0,6.0)
 \resizebox{6cm}{!}{
\psfig{file=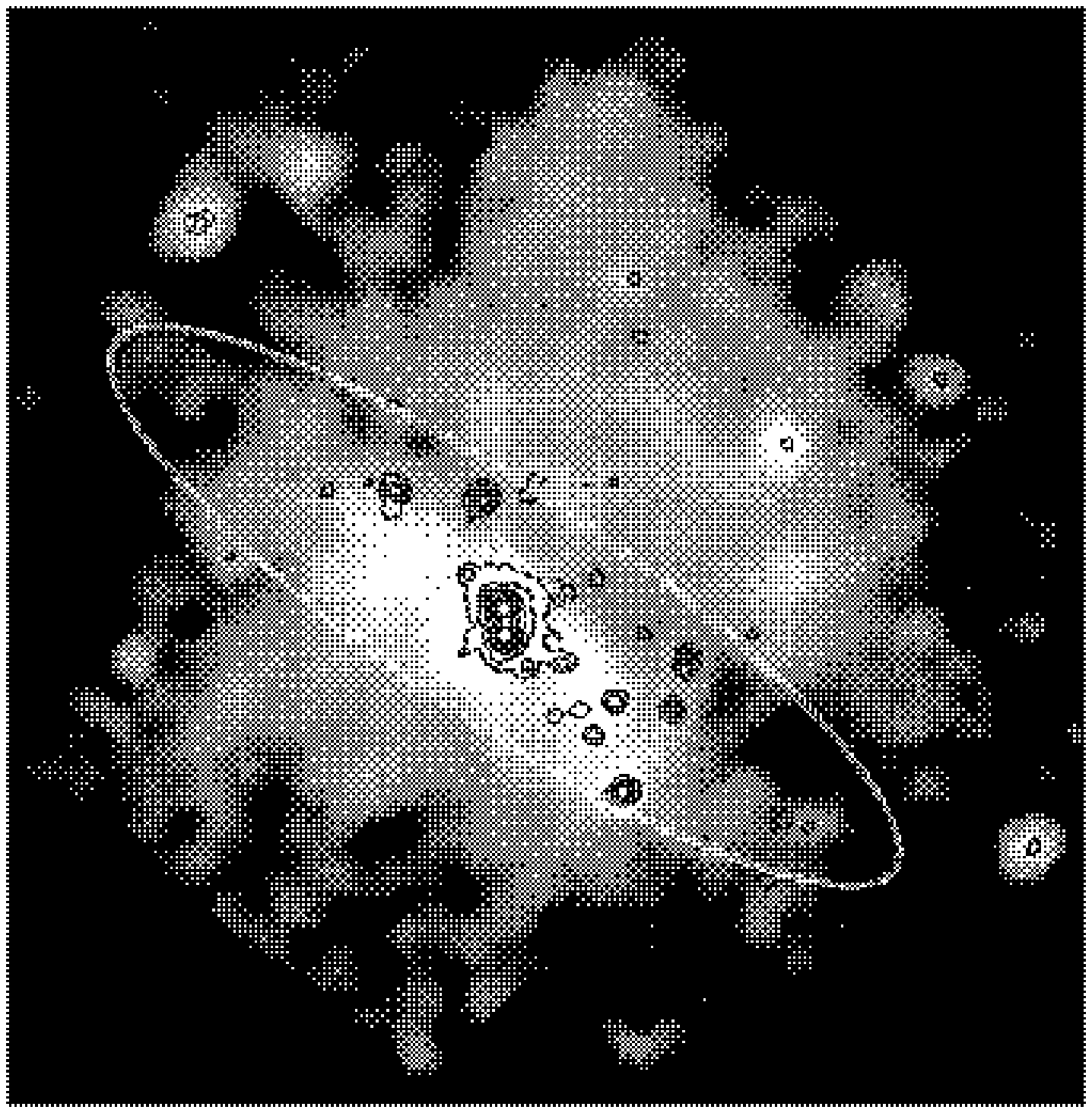,width=6cm,clip=}}
 \end{picture}}
\put(6.0,2.6){
\parbox[]{55mm}
{In this figure, the X-ray halo of NGC~253 in the soft (0.2-0.5 keV)
energy band  observed with XMM-Newton (courtesy of W. Pietsch) is shown,
with the
hard emission overlayed as contours. The oval shape indicates the
approximate size and orientation of the optical disk.  A few point
sources are evident in the halo region, otherwise the area above and
below the plane appears uniformly filled at these energies. 
}}
\caption[]{}
\label{n253}
\end{picture}
\end{figure*}

\section{Characteristics of the ISM in starburst galaxies/regions}

Starburst galaxies are the subject of separate talks in this conference
(cf. talks by Dahlem and by Dettmar), so I will briefly review the X-ray
evidence, mostly at large scale.  The data are fast accumulating but
they have not been properly digested yet, and most of the results are
still rather qualitative and based on a few examples only.

\begin{figure}
\hskip 0.4cm
\resizebox{12cm}{!}{
\psfig{file=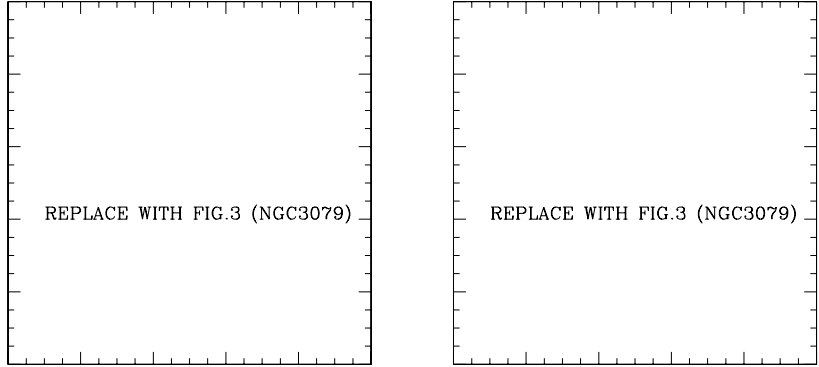,width=4.5cm}
\psfig{file=boxfig3.ps,width=4.5cm}
 }
\caption{XMM-Newton data of NGC~3079 in the soft band indicate a large
scale emission above/below the plane of the galaxy, as defined by the
optical image.  The filamentary structure of the emission is evident. 
Filaments are traced out to 13-17 kpc.  Three main spurs plus a smaller
one are responsible for the X-shaped morphology of the halo.
From Breitschwerdt et al. 2002.
}
\label{n3079}
\end{figure}

A large fraction of the total emission of starburst galaxies 
in the soft X-ray band is due to the
hot ISM. 
The fraction varies in different galaxies, with $>$ 20\% in
NGC~3256 (Moran et al. 1999; Lira et al 2002), to $\sim$50\% in the
Antennae \cite{fabbianoetal2001}, to $\sim$80\% in M82 \cite{zezas}.
Hot gas is found mainly in three regions: (A) the large scale
distribution 
is associated with extraplanar emission, in the galaxy's \underline{\bf halo},
usually associated with evidence of halo emission at
other wavelengths (HI, radio
continuum, H$\alpha$ outflows). The emission can be traced to quite
large distances: the rather spectacular halo of NGC~253 observed with
XMM-Newton in the very soft energy band
(Fig.~\ref{n253}) fills a very large region above and below the plane.
The halo is usually fueled/fed by a nuclear/galactic
outflow/wind.  Often ``horns''  or X-shaped morphologies are observed in
galaxies viewed edge-on.   
A much finer filamentary structure is also observed in halos, e.g. 
in NGC~3079's halo \cite{breit} with XMM-Newton data  and 
in NGC~253 itself with Chandra data, with spatial fluctuations of $\sim$130
pc in size \cite{stricketal02}.
In both cases, these point to a close relation
with similar structures in the line emission (cf. above and 
Fig.~\ref{strick}).  (B) The \underline{\bf disk} appears to contain a
multi-phase hot ISM.  To use the case of NGC~253 again, the data require    
two temperatures (0.15; 0.5 kT) with no additional absorption 
and show  unquestionable
signs of plasma lines, e.g. O{\sc V}{\sc ii}, Fe{\sc X}{\sc V}{\sc
ii},  that point to emission from a hot
ionized medium \cite{pietsch}.  
It is possible that the multi-temperature of the gas in the disk  is to
be attributed in part to contamination from  emission
at higher galactic latitudes projected onto the disk. (C)  Possibly the
stronger emission is associated with the
\underline{\bf nuclear/circumnuclear} regions, from which  plumes and
outflows are likely to be fueled.   These regions appear to be hotter
than the rest of the disk (e.g. kT$\sim$6 keV is found in the inner
regions of NGC~253 [again!], \citeauthor{pietsch}\citeyear{pietsch}), and might contain a
higher metal fraction than the outer regions.

A much larger and better studied sample of objects needs to be examined
for more quantitative analysis and a broader range of parameters need to
be explored before a classification of the X-ray properties of the hot
ISM in these galaxies can be finalized.

\section{Early type galaxies and groups}

The large quantities of gas observed in this class of sources have now
been studied in some detail and several general characteristics of the
large scale emission are now
relatively well established.  However, little is known of the emission
at intermediate to small scales.  New observations have now shown that 
the gas distribution in the inner regions of early type galaxies can be  
quite spectacular.  At small radii the emission is often disturbed, and 
arcs-tails-peculiar/unexpected  shapes are coupled with inhomogeneous
spectral characteristics of the gas.   I will concentrate on a few
examples from the literature that will illustrate how gas distribution
in early type galaxies can be just as interesting as in later types, 
and rather more unexpected! 

\begin{figure}
\psfig{file=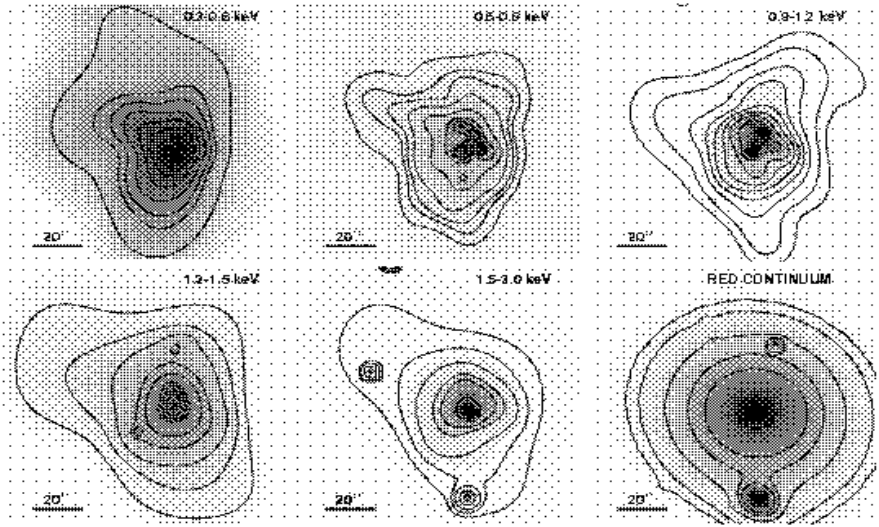,width=11.7cm}
\caption[]{Adaptively smoothed X-ray data of the central region of 
NGC~5846 at different energies, as
indicated in each panel.  The red continuum image of the same region
(data from the ESO 3.6m + EFOSC) is
shown in the bottom right cannel. The comparison between the different
images is a clear example of the complexity of the X-ray morphology,
and shows how little it reproduces the stellar light distribution. 
}
\label{n5846x}
\end{figure}

\subsection{NGC~5846}

This galaxy is at the center of a small group, and optically is a rather
unperturbed object.  The X-ray emission is very extended, roughly
azimuthally symmetric and regular, with a large
total X-ray luminosity of L$_X \rm (0.2-2 keV)$ 7 $\times 10^{41}$ erg s$^{-1}$
within r=10$'$ \cite{finoguenovetal}.
The emission on the sub-arcminute scale instead is 
much more spectacular and complex: as shown by Fig.~\ref{n5846x}, an
arc structure (``hook"), a tail pointing S and several knots are
evident in the soft energy band, and their relative strengths 
are clearly energy-dependent.

There is also a curious correspondence with the optical line emission
and with the dust distribution, both on scales of
$1'-2'$, as already
discovered with the ROSAT HRI \cite{trinchierietal97}, and on
much smaller scales of $2''- 4''$ (300-600 kpc) discovered with Chandra
\cite{trinchierigoud}.
The correspondence is extremely good,  but the physical
link between the two emissions is still rather puzzling: 

\smallskip
\noindent\hangindent 10pt $-$ there is no nuclear activity nor a  strong Radio Source in the
center that would justify/trigger gas excitation 

\noindent\hangindent 10pt $-$ the origin of the two gas phases are not the same: \\
$\bullet$ H$\alpha$ is most likely of ''external'' origin, acquired from 
merging or interaction with smaller, gas rich bodies together with dust.  
$Here$ the internal kinematics supports external origin\\
$\bullet$ X rays  are ``internal'', from the ISM of the galaxy
presumably accumulated from stellar mass loss.

\begin{figure}
\hskip -0.5cm
\resizebox{12cm}{!}{
\psfig{figure=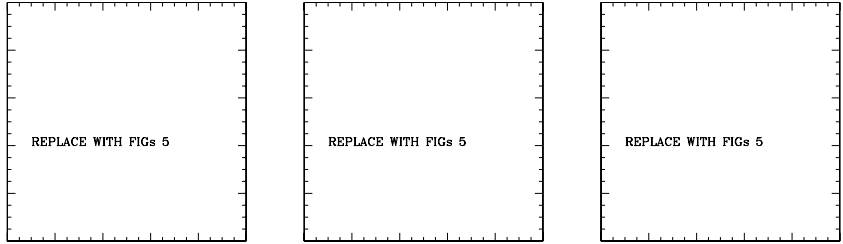,width=3.0cm}
\psfig{figure=boxfig5.ps,width=3.0cm}
\psfig{figure=boxfig5.ps,width=3.0cm}
 }
\caption[]{H$\alpha$+[N{\sc ii}] image of the inner regions of
NGC~5846 (LEFT) on the same scale as the 0.2-2.0 keV emission observed
with Chandra (MIDDLE).  The H$\alpha$+[N{\sc ii}] contours on the
X-ray smoothed image (RIGHT) indicate that both morphology and scales
are very closely related in this region (from \citeauthor{trinchierigoud}
[\citeyear{trinchierigoud}]).
}
\label{n5846xha}
\end{figure}

\smallskip
\noindent\underline{Why are their morphologies so similar} then 
(Fig~\ref{n5846xha})?
A possible link between H$\alpha$ [+dust] and X rays is through  thermal
conduction \cite{cowie,sparksetal,dejong}.  Cold gas+dust are  acquired
from the outside at the same time; dust/cold gas act simultaneously on
the hot coronal gas, inducing local cooling + excitation into emission;
this produces an expected L$\rm _{H\alpha}$  $\sim 1.5 \times 10^{39}$
T$^{3/2}_7$ n$_{0.01}$ erg s$^{-1}$ \cite{cowie}.  For the temperature
and densities derived from the X-ray observation e.g. in the ``hook",
of T$\sim 6 \times 10^6$ K and $n \sim 0.35$ cm$^{-3}$ respectively,
the H$\alpha$ luminosity should be $\sim 2.4 \times  10^{40}$ erg
s$^{-1}$, to be compared to a measured L$\rm _{H\alpha}$ $\sim 1.3
\times 10^{40}$ erg s$^{-1}$.  {\bf Too good to be true!}

Can we then assume that this mechanism is responsible for the complex
morphologies at small radii seen in other galaxies?  The number of cases
with both good X-ray and H$\alpha$ data is limited, but there are a few
examples -- unfortunately so far NGC~5846 remains the best and perhaps
only example for which such close link is observed. For other objects
there seem to be a need of other explanations for the small scale
morphological perturbation, and unfortunately again each object appears 
so far to be ``unique" (but the cases studies are very few!)

\begin{figure*}
\resizebox{12cm}{!}{
\psfig{figure=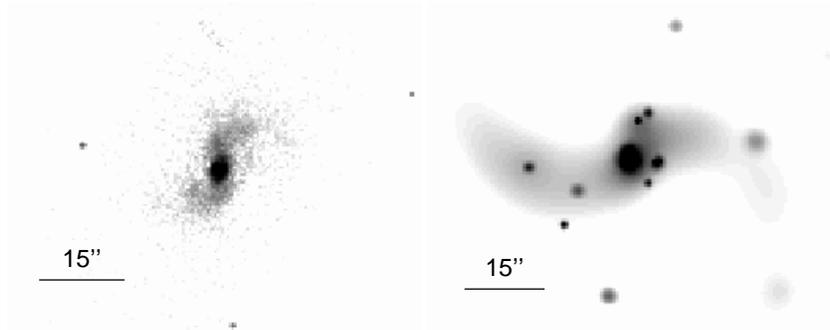,width=4.0cm}
 }
\caption[]{Inner regions on NGC~1553 in 
H$\alpha$+[N{\sc ii}] (left; ESO NTT) and 0.3-1.0 keV band (Chandra
archive).  In the very soft X-ray band
a spiral structure roughly symmetric about a central point-like source
is evident, with a possible NS oriented, 10$''$ long, inner bar,
reminiscent of the H$\alpha$+[N{\sc ii}] structure.
No such feature is evident in the optical continuum or
the X rays above 1 keV.  
}
\label{n1553}
\end{figure*}

\subsection{NGC~1553}

The overall emission from NGC~1553 indicates an 
extended component, slightly elongated NW-SE,
consistent with the optical axis \cite{trinchierietal97}. 
Even with the lower resolution ROSAT data, several features
could be identified that indicated structures at smaller, arcmin size,
angular scales. 
New Chandra data now show that the inner regions are quite complex 
\cite{blantonetal}:
at energies below 1 keV there is a clear evidence of a
spiral structure, a possible inner bar, and a twisted NW-SE elongation
at larger radii (cf. \citeauthor{blantonetal} and Fig.~\ref{n1553}).
At intermediate radii, ROSAT HRI  data had already
indicated an overall similarity with line emission, although not as
close as in the NGC~5846 case \cite{trinchierietal97}.   This similarity appears
to break down at small radii \cite{blantonetal}: there could
be a common NS ``bar" in the inner 10$''$ (\cite{trinchierietal97}
\cite{rampazzo}, but it is hard to envision 
thermal conduction to work along the spiral structure 
where the H$\alpha$ does not extend. 
New high resolution Fabry-Perot interferometric data 
have further shown quite complex gas dynamics in NGC~1553 \cite{rampazzo}, 
making the possible similarity with the hot phase more
difficult to interpret.
The origin of the spiral feature is better interpreted  with adiabatic or
shock compression of the ambient gas possibly due to interaction with the radio source
\cite{blantonetal}.

\subsection{NGC~4636}

\begin{figure*}
\unitlength1.0cm
\begin{picture}(12,4.2)
\thicklines
\put(-0.3,0){
\begin{picture}(5.5,4.2)
\psfig{figure=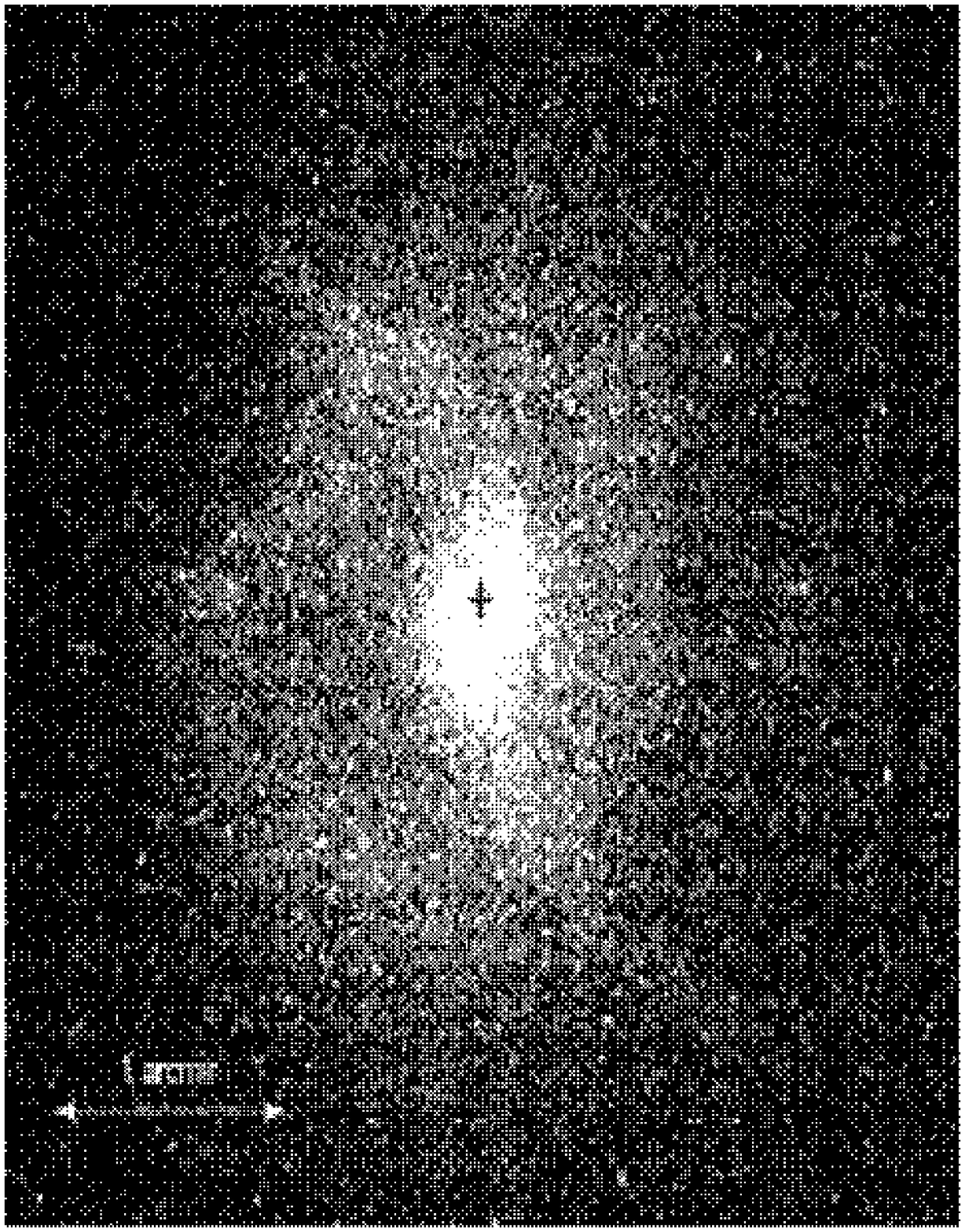,width=5.0cm,height=4.2cm}
 \end{picture}}
\put(5.0,1.8){
\parbox[]{65mm}
{Right (from \citeauthor{jonesetal} [\citeyear{jonesetal}]): 
Full resolution ACIS-S image of the central regions of NGC 4636 
in the 0.5-2.0 keV energy range.  A plus sign marks the galaxy center.
\label{n4636}
}}
\caption[]{}
\end{picture}
\end{figure*}

A wealth of X-ray data has been accumulated for this X-ray bright
galaxy since $Einstein$ Observatory times, that indicate a
very extended distribution of the X-ray emission, due almost entirely 
to hot gas
at an average temperature of $\sim 1$ keV and total luminosity of 
L$_X \sim 2 \times 10^{41}$ erg s$^{-1}$ 
\cite{trinchierietal94, matsushita, buote}.
From an $Einstein$ HRI observation, \cite{stanger}
had already proposed an asymmetric gas distribution at
small radii.  Chandra data now clearly show symmetric, 8 kpc long,
clearly defined spiral-like twisted structures in  the 
inner few arcmin region of this galaxy (Fig.~\ref{n4636},
\cite{jonesetal}), coupled
with a complex temperature structure of the gas,   
a weak but measurable temperature gradient (but no cooling flow!)
in the inner regions and a number of discrete spectral lines measured
with XMM-Newton data \cite{xuetal}.  To date there are no reports  
of a close  morphological relation between the hot and the
warm phases, although H$\alpha$ is detected in this object
\cite{demoulin} with chaotic gas kinematics and with 
evidence for a
kinematically distinct inner region like NGC~5846 \cite{caonetal}.
\citeauthor{jonesetal}  propose that the X-ray 
morphology is formed as a result of shocks driven by a nuclear outburst
in the recent past.  These outbursts would also have implications both
in the accumulation of material in the galaxy's center, and in the
fueling of a central AGN.

\section{Unexpected phenomena}

Another set of surprises come from the presence of ``unsuspected'' large
scale shocks, not seen/foreseen from the optical data.  How many are we
missing because we are not looking at the right wavelength?  

\subsection{IC1262}

This galaxy, brightest in a small poor group,  became interesting because
of its high $\rm L_x/L_b$ ratio as seen in the ROSAT All Sky Survey.  
X-ray emission was later measured to be very extended ($>$400 kpc), 
suggesting a large contribution from  the 
group's potential \cite{trinchieripietsch}.
However, more interesting, the HRI saw
an  ``arc'' on $\le$arcmin scales, that was interpreted as the signature
of a shock.

\begin{figure}
\resizebox{11cm}{!}{
\psfig{file=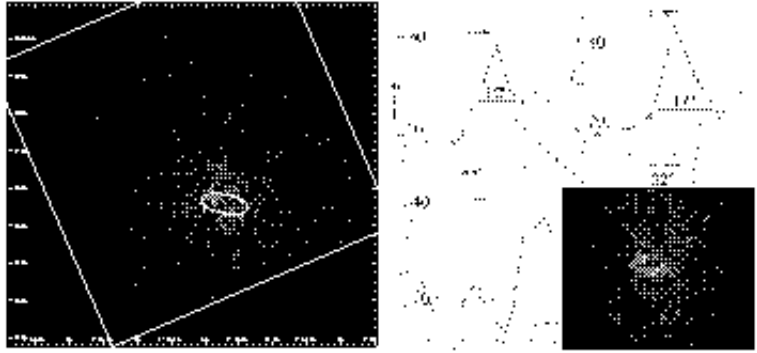}
}
\caption{The full Chandra field of IC1262 (LEFT).   The oval
indicates the optical position of the galxy. 
A turmoil at the center of the more diffuse emission 
and a sharp 
discontinuity in surface brightness  are evident even
in the raw data, in cuts across the narrow features visible. 
}
\label{ic1262}
\end{figure}

New Chandra data have now shown a quite dramatic view of the system
(Fig.~\ref{ic1262}):  a sharp discontinuity east of the central galaxy,
with steep drops and a relatively narrow feature along a possible shock
front, plus an arc to the NNW,  are all indicative of a turmoil in the
high energy component,  possibly a trace of shocked material caused
either by peculiar motions in the system or by a recent merger
process.  Although a better sampling of the velocity field is needed,
neither the optical classification of the galaxy as a cD nor the small
velocity dispersion for the group (300 km s$^{-1}$, \cite{wegner})
could have suggested anomalous motions of this nature.  A proper
assessment of the X-ray properties in this and similar systems might be
crucial for a more complete understanding of the dynamical and
evolutionary properties of small galaxy systems.

\section{What can we conclude from all of this?}

It is probably too early to draw conclusions from the evidence
accumulated so far.  However, it is becoming clearer that the small
scale gas morphology indicates that the ``early- type'' galaxy
population is far from  ``quiescent''. Same can be said for ``cD''
galaxies in small/sparse groups.   This would indicate that there is
ground for more detailed studies of the high energy emission in
galaxies; these might reveal unsuspected phenomena like shocks that are
crucial for our understanding of the evolution of these systems.  The
multiwavelength approach is the only one that can provide a complete
picture of all the forces at play, and the study of the gaseous
components, though complex, has a very high potential for easy
discovery of extreme, but perhaps not uncommon, phenomena.

\end{article}
\end{document}